\documentclass[twoside]{article}
\usepackage{fleqn,espcrc2}

\usepackage{graphicx}
\usepackage[figuresright]{rotating}

\newcommand{\AmS}{{\protect\the\textfont2
  A\kern-.1667em\lower.5ex\hbox{M}\kern-.125emS}}

\title{On the Thermodynamics of a Heavy Quark-Antiquark Pair}

\author{D. Antonov, S. Domdey, H.-J. Pirner
\address{Institut f\"ur Theoretische Physik, Universit\"at Heidelberg,\\
 Philosophenweg 19, D-69120 Heidelberg, Germany}}

\begin{document}

\begin{abstract}
Thermodynamics of a heavy quark-antiquark pair in SU(3)-QCD is studied
  below the deconfinement critical temperature, $T_c$.
  In the quenched case, a model of the string passing through heavy
  valence gluons yields a correct estimate of $T_c$ and a behavior of the string tension near $T_c$.
For two light flavors, entropy and internal energy can be obtained from the partition function of heavy-light 
  mesons and baryons. They are in a good qualitative agreement with the lattice results. 
\vspace{1pc}
\end{abstract}

\maketitle

\section{INTRODUCTION}
In the last few years, various indications appeared that the quark-gluon plasma is not a weakly coupled 
system even up to temperatures of the order of a few times the deconfinement one, $T_c$. For instance, RHIC data 
point to a nearly perfect-fluid type behavior of the plasma at such temperatures, which is characterized by the shear viscosity to the entropy-density ratio of the order of 0.1, whereas the values for this ratio predicted by perturbative QCD are larger than 1. 

Besides experimental data, a lot of information on the nonperturbative 
properties of the plasma comes from the lattice. Among the recent results obtained there and calling for a theoretical explanation,
are the anomalously large (from the perturbation theory standpoint) maxima of the entropy and internal energy of the static quark-antiquark pair in unquenched QCD around $T_c$~\cite{latint1}, \cite{tc}. This work is aimed at a description of these data below $T_c$. The strategy of our analysis is the following. In the next section, we will determine $T_c$ and the effective 
string tension $\sigma(T)$ in the quenched SU($N_c$) QCD within the so-called gluon-chain model. In section~3, 
with the use of $\sigma(T)$ adjusted to the unquenched ($N_c=3, N_f=2$)-case, we will calculate, within the relativistic quark model, the partition function of heavy-light mesons and baryons. Entropy and internal energy stemming 
from this partition function will further be compared to the corresponding lattice data, and in the Summary section 
conclusions will be drawn.

\section{STATIC $Q\bar Q$-PAIR WITH GLUONS}
It is generally accepted that the $Q\bar Q$-string, which sweeps out the flat surface of the corresponding Wilson loop, 
is produced by soft (stochastic background) gluonic fields. On top of those, fluctuations of the gauge field exist, which lead to
string vibrations. These fluctuations are called valence gluons, and the $Q\bar Q$-string passing through one such gluon is called hybrid. At asymptotically large $Q\bar Q$-separations of interest, one has a string passing through numerous valence gluons, which we will call gluon chain. Further, when one starts heating the system, at low temperatures the free energy of one string bit between two nearest gluons in the chain, being a constant, exceeds thermal mass of a gluon, which grows with $T$ linearly from zero on. The motion of gluons at low temperatures is therefore guided by the string -- they
move collectively and do not affect string's dynamics. However, at a certain temperature $T_0$ smaller than $T_c$,
the gluon's thermal mass becomes larger than the free energy of one string bit. From this temperature on, the system looks totally different, since gluons from the string's standpoint are now (nearly) static. 
Therefore, at $T_0<T<T_c$, a gluon chain is nothing but a sequence of 
static nodes with adjoint charges, connected by independently fluctuating string bits. At the moment of formation of such a chain, 
its end-point originating from the heavy $Q$ performs a random walk towards $\bar Q$ over the lattice of static nodes.
The entropy of such a random walk turns out to be large, namely proportional to its length, and eventually leads to the deconfinement phase transition in this model. The reason is 
that color may alter from one node to another, i.e. every string bit may transport each of the $N_c$ colors. The total number of 
states of the gluon chain is therefore $N_c^{L/a}$, where $L$ is its length and $a$ is the length of one bit.

Taking into account that the full free energy is the sum of the usual linear potential and the 
free energy of the random walk, we have for the effective string tension:
$$
\sigma(T)=\sigma-T\ln\frac{{\cal Z}(R,T)}{{\cal Z}(R,T_0)}\Biggr|_{R\to\infty},$$
where 
$${\cal Z}(R,T)=\sum\limits_{n=-\infty}^{+\infty}\int_0^\infty\frac{ds}{(4\pi s)^2}\times$$
$$\times\exp\left[-\frac{R^2+(\beta n)^2}{4s}-
\frac{s}{a}\left(\beta\sigma-\frac{\ln N_c}{a}\right)\right]$$
is the partition function of the random walk. Here, $\beta\equiv1/T$, $\sigma=(440{\,}{\rm MeV})^2$ is the zero-temperature string tension, $s=aL$ is the Schwinger proper time, and $n$ is the number of a Matsubara mode. This formula takes into account the confining potential between the end-point of the walk and its starting point (chosen at the origin), as well as the 
entropy factor discussed above. At asymptotically large $R$'s of interest, only the contribution of the ($n=0$)-term is essential, and we obtain 
$$
\sigma(T)=\sigma+T\Biggl[\sqrt{\frac{\sigma\beta}{a}\left(1-T\frac{\ln N_c}{\sigma a}\right)}-$$
\begin{equation}
\label{1}
-\sqrt{\frac{\sigma\beta_0}{a}\left(1-T_0\frac{\ln N_c}{\sigma a}\right)}\Biggr].
\end{equation}
An estimate for $T_c$ stems from the condition that the argument of the first square root vanishes:
\begin{equation}
\label{2}
T_c\Bigr|_{N_c>1}=\frac{\sigma a}{\ln N_c}.
\end{equation}
Equating $T_c$ to the modern $N_c=3$ lattice value~\cite{latint1}, 270 MeV, we obtain an effective length of one string bit 
$a\simeq0.31{\,}{\rm fm}$. This is larger than the minimal possible value of this quantity, $a=0.22{\,}{\rm fm}$ -- the 
so-called vacuum correlation length~\cite{a}, which defines the onset of a string-bit formation. The temperature $T_0$, 
below which the lattice of valence gluons does not exist, can be defined from the condition $\sigma(T_c)=0$, which yields
$$
T_0=\frac{T_c}{\ln N_c+1}\simeq130{\,}{\rm MeV}.$$
An important finding of this model is the behavior 
$$\sigma(T)\sim\sqrt{T_c-T}~~~ {\rm at}~~ T\to T_c.$$ 
It is the same as the one which follows from the 
Nambu-Goto model for the two-point correlation function of Polyakov loops~\cite{pa}. 

Let us finally consider the limiting case when string bits cannot alter color, i.e.
one should formally set in Eq.~(\ref{1}) $N_c=1$, that yields
$$\sigma(T)=\sigma+\sqrt{\frac{\sigma T}{a}}\left(1-\sqrt{\frac{T}{T_0}}\right).$$
Determining in this case $T_0$ directly from the equality of the gluon's thermal mass in QCD 
to the free energy of one string bit, we have
$$T_0=\frac{\sigma a}{g},~ T_c=\frac{\sigma a}{4g}\Bigl(1+\sqrt{1+4\sqrt{g}}\Bigr)^2.$$
Setting for an estimate $a=0.22{\,}{\rm fm}$, $g=2.5$, we obtain still reasonable values $T_c=290{\,}{\rm MeV}$, 
$T_0=85{\,}{\rm MeV}$. However, the fundamental difference of this limiting case from the realistic one, $N_c>1$,
is that here 
$\sigma(T)\to\frac{1}{2}\sqrt{\frac{\sigma(1+4\sqrt{g})}{aT_c}}(T_c-T)\sim(T_c-T)$ 
at $T\to T_c$. Such type of the critical behavior characterizes the 2d Ising model, which
by no means can be realized in the 4d quenched SU($N_c$) QCD. 
The same linear fall-off of $\sigma(T)$ with $(T_c-T)$ one finds also 
in the Hagedorn phase transition and in the deconfinement scenario based on the condensation of long closed strings~\cite{1}.
It is a mere consequence of the formula $\sigma(T)=\sigma-\frac{TS}{R}$ at $S\propto R$.

\section{STATIC $Q\bar Q$-PAIR WITH LIGHT QUARKS}
Let us now consider the unquenched case, where, at a certain distance, the $Q\bar Q$-string breaks due to the production of 
a light $q\bar q$-pair. The subsequent hadronization process leads to the formation of 
heavy-light mesons ($\bar Qq$) and heavy-light-light baryons ($Qqq$), as well as their antiparticles. We will consider the ($N_f=2$)-case, with 
light $u$- and $d$-quarks, and use the value $T_c=200{\,}{\rm MeV}$~\cite{tc}. Equation~(\ref{2}) then yields an effective 
value of $a=0.23{\,}{\rm fm}$, which should now be used in Eq.~(\ref{1}) to determine $\sigma(T)$. Then, for a heavy-light meson, the Hamiltonian of the relativistic quark model reads
\begin{equation}
\label{H}
H_{\bar Qq}=m_{\bar Q}+\sqrt{{\bf p}^2+m_q^2}+V(r).
\end{equation}
Here, $V(r)=\sigma(T)r-(2-\delta)\sqrt{\sigma(T)}$, $m_{\bar Q}$ is the heavy-antiquark mass, whereas $m_q$ is the constituent mass of a light quark, $m_q\simeq 300{\,}{\rm MeV}$. 
The subtraction of $2\sqrt{\sigma(T)}$ in $V(r)$ is known to be important to reach an agreement between the predictions of the relativistic quark model to the phenomenology of meson spectroscopy~\cite{const1}, \cite{const2}. 
An additional correction, $\delta\sqrt{\sigma(T)}$, is needed to obey the normalization condition
$F\to2(m_{D^0}-m_c)$ at $T\to 0$. Here, $F$ is the resulting free energy of two noninteracting mesons, 
$m_{D^0}=1.864{\,}{\rm GeV}$ is the $D^0$-meson mass, $m_c=1.48{\,}{\rm GeV}$ is the $c$-quark mass.
Further, the square root in the relativistic kinetic energy
can be disentangled by using the integration over an auxiliary parameter:
$$\exp\left(-\beta \sqrt{{\bf p}^2+m_q^2}\right)=\vspace{-1.5mm}$$ 
$$=\frac{2}{\sqrt{\pi}}
 \int_0^\infty d\mu \exp\left(-\mu^2-\frac{\beta^2}{4\mu^2}\left({\bf p}^2+m_q^2\right)\right).$$
One then arrives at a 3d Schr\"odinger equation of the form
$$
(K\partial^2+M|{\bf x}|)\psi({\bf x})=E\psi({\bf x}),$$
where $K$ and $M$ are positive constants of dimension $[{\rm length}]$ and $[{\rm mass}]^2$, respectively. 
Its eigenenergies have the form~\cite{const2}
$$E_{n_r,l}= \alpha_{n_r,l}\, (K M^2)^{1/3},$$
where $\alpha_{n_r,l}$'s are positive numbers with $\alpha_{00}\simeq 2.34$. Further strategy for the calculation of the 
partition function of a meson~\cite{const2} consists of accounting for its ground state exactly and modeling its higher 
eigenenergies by those of a 3d harmonic oscillator with the frequency 
$$\omega\equiv(K M^2)^{1/3}=\left(\frac{\beta\sigma^2(T)}{4\mu^2}\right)^{1/3}.$$
Then, because of the 
degeneracy factor $(n/2+1)(n+1)$ for the oscillator, we have
$$
{\cal Z}_{\bar Qq}={2\over \sqrt{\pi}} \exp\left(-\beta m_{\bar Q} + 2\beta \sqrt{\sigma(T)}\right)\times\vspace{-1mm}$$ 
$$\times\sum_{n=0}^\infty
\left(\frac{n}{2}+1\right)(n+1)\int_0^\infty d\mu\times\vspace{-0.7mm}$$
$$\times\exp\left[-(n+\alpha_{00})\beta\omega-\mu^2-\frac{\beta^2 m_q^2}{4\mu^2}\right].$$
The free energy of two noninteracting mesons reads $F=-T\ln(4{\cal Z}_{\bar Qq}^2)$, where the factor 4 is a product of two
flavors and two spin states of the mesons (which are antiparticles of each other).
Doing the sum over $n$ analytically, we arrive at the following result:
$$
F=-T \ln\frac{16}{\pi}- 4\sqrt{\sigma(T)}
-2 T \ln\Biggl[\int_0^\infty d\mu\times\vspace{-1mm}$$
\begin{equation}
\label{fe}
\times \exp\left(-\mu^2 - 
\frac{\beta^2 m_q^2}{4 \mu^2}\right) \frac{\exp(-\alpha_{00}\beta\omega)}{(1-\exp(-\beta\omega))^3}\Biggr].
\end{equation}
The remaining $\mu$-integration has been done numerically. The resulting value of the parameter $\delta$ is 0.344.
Entropy and internal energy can further be calculated by the standard formulae $S=-\partial F/\partial T$, 
$U=F+TS$. Note that, when calculating the entropy, the derivative $\partial/\partial T$ 
should not act on the Hamiltonian in the partition function.

Similarly, one can treat $(Qqq)$-baryons, whose Hamiltonian reads
$$H_{Qqq}=m_Q+\sum\limits_{i=1}^{2}\left(\sqrt{{\bf p}_i^2+m_q^2}+V(r_i)\right).$$
Here, we have approximated the position of the baryon string-junction point by the position of $Q$, 
that is legitimate due to the heaviness of $Q$. The value of the parameter $\delta$, at which the free energy of a baryon goes
to $m(\Lambda_c^+)-m_c$ at $T\to 0$, is 0.393. Here, $m(\Lambda_c^+)=2.286{\,}{\rm GeV}$ is the mass of the $\Lambda_c^+$-baryon. In the figures,
we plot the entropy and the internal energy of mesons together with baryons, at the averaged value $\delta=0.37$. 
In the same figures, we plot the corresponding lattice data 
for the $Q\bar Q$-pair~\cite{tc}.

\begin{figure}
\includegraphics[clip,scale=0.85]{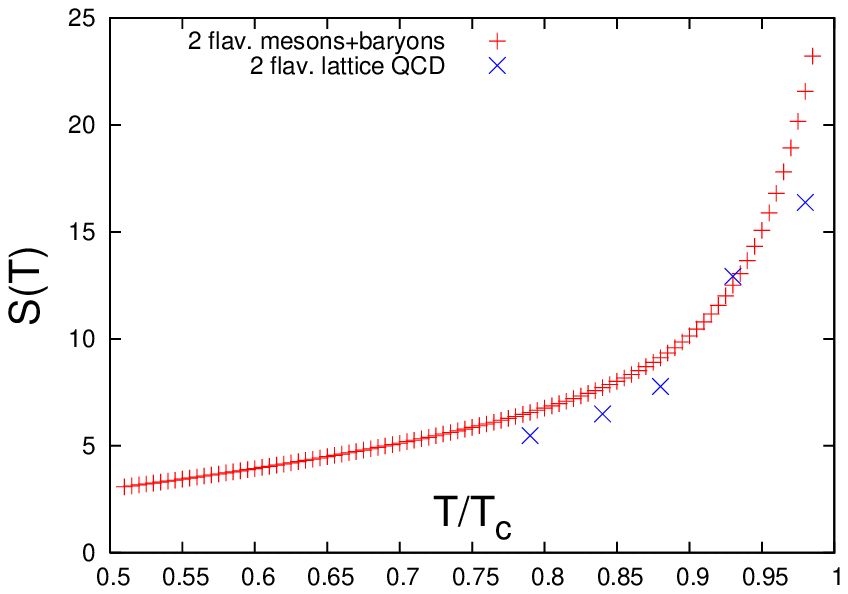}%
\end{figure}

\begin{figure}
\includegraphics[clip,scale=0.85]{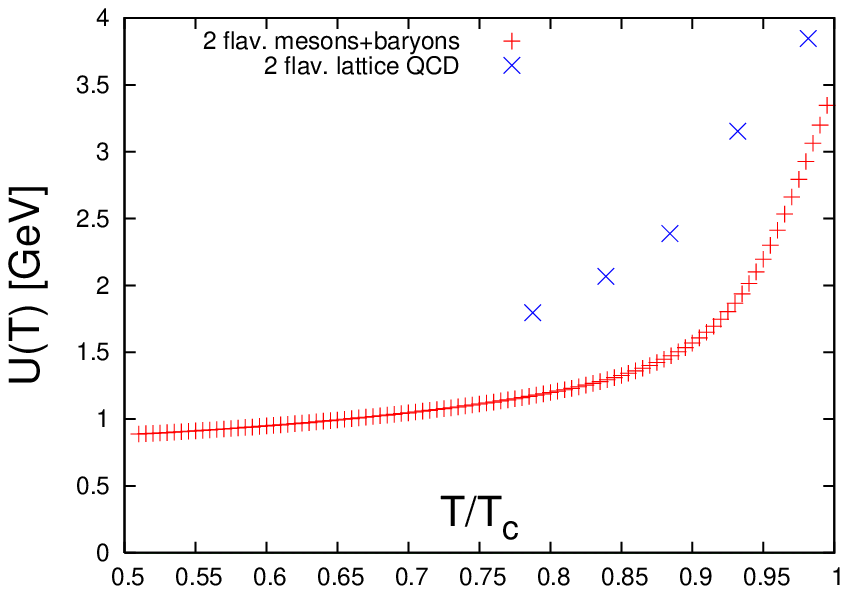}%
\end{figure}

\section{SUMMARY}
In this work, we have analytically addressed thermodynamics of the static $Q\bar Q$-pair below the deconfinement phase transition, in quenched and unquenched QCD. In the quenched SU($N_c$)-case, within the so-called gluon-chain model, 
we have derived an effective temperature-dependent string tension. With this string tension adjusted to the unquenched ($N_c=3, N_f=2$)-case, we have calculated 
the entropies and internal energies of heavy-light mesons and baryons. 
In general, the results obtained are in a good agreement with the corresponding recent lattice data for the same quantities. Our analysis shows also that the entropy and the 
internal energy of heavy-light mesons alone are not enough to reproduce correctly the lattice data.

\section*{ACKNOWLEDGMENTS}
D.A. acknowledges the organizers of the conference 'QCD 06' (Montpellier, France, 3-7 July, 2006)
for an opportunity to present these results in a stimulating atmosphere. The work of D.A. has been supported through 
the contract MEIF-CT-2005-024196. S.D. thanks P.~Petreczky and F.~Zantow for providing him with the details of the lattice data.


\begin{thebibliography}{50}

\bibitem{latint1}
P.~Petreczky and K.~Petrov,
Phys.\ Rev.\ D 70 (2004) 054503; P.~Petreczky,
Eur.\ Phys.\ J.\ C 43 (2005) 51.

\bibitem{tc}
O.~Kaczmarek and F.~Zantow,
``Static quark anti-quark interactions at zero and finite temperature QCD.
II: Quark anti-quark internal energy and entropy,''
preprint hep-lat/0506019 (unpublished).
  

\bibitem{a}
A.~Di~Giacomo and H.~Panagopoulos, Phys.\ Lett.\ B 285 (1992) 133.

\bibitem{pa}
R.~D.~Pisarski and O.~Alvarez,
Phys.\ Rev.\ D 26 (1982) 3735.

\bibitem{1}
See, e.g., H.~Meyer-Ortmanns,
Rev.\ Mod.\ Phys.\ 68 (1996) 473.

\bibitem{const1}
D.~Gromes,
``Theoretical understanding of quark forces,'' 
preprint HD-THEP-89-17 (unpublished).

\bibitem{const2} 
H.~J.~Pirner and M.~Wachs, Nucl.\ Phys.\ A 617 (1997) 395.




\end{thebibliography}
\end{document}